# Predicting Distant Metastases in Soft-Tissue Sarcomas from PET-CT scans using Constrained Hierarchical Multi-Modality Feature Learning

Yige Peng, Lei Bi, *Member, IEEE*, Ashnil Kumar, *Member, IEEE*, Michael Fulham, Dagan Feng, *Fellow, IEEE*, and Jinman Kim, *Member, IEEE*

*Abstract*— **Distant metastases (DM) refer to the dissemination of tumors, usually, beyond the organ where the tumor originated. They are the leading cause of death in patients with soft-tissue sarcomas (STSs). Positron emission tomography-computed tomography (PET-CT) is regarded as the imaging modality of choice for the management of STSs. It is difficult to determine from imaging studies which STS patients will develop metastases. 'Radiomics' refers to the extraction and analysis of quantitative features from medical images and it has been employed to help identify such tumors. The state-of-the-art in radiomics is based on convolutional neural networks (CNNs). Most CNNs are designed for single-modality imaging data (CT or PET alone) and do not exploit the information embedded in PET-CT where there is a combination of an anatomical and functional imaging modality. Furthermore, most radiomic methods rely on manual input from imaging specialists for tumor delineation, definition and selection of radiomic features. This approach, however, may not be scalable to tumors with complex boundaries and where there are multiple other sites of disease. We outline a new 3D CNN to help predict DM in STS patients from PET-CT data. The 3D CNN uses a constrained feature learning module and a hierarchical multi-modality feature learning module that leverages the complementary information from the modalities to focus on semantically important regions. Our results on a public PET-CT dataset of STS patients show that multi-modal information improves the ability to identify those patients who develop DM. Further our method outperformed all other related state-of-the-art methods.**

## I. INTRODUCTION

SOFT-TISSUE sarcomas (STSs) are a heterogeneous group of malignant tumors that arise in connective tissue in the body [1]. About 25% of STSs patients develop distant metastases (DM) that are the main cause of death [2], [3]. DM refers to the spread of tumors from the site of origin, usually, to other structures and viscera in the body; tumors can spread to other locations in the same organ but these tend to produce local problems. For high-grade STSs, 50% of STS patients will develop DM and the median survival is 11.6 months after they develop DM [3]. It is hoped that the early identification of patients at a high risk of developing DM will enable the introduction of more effective therapies [4], [5]. Positron emission tomography-computed tomography, using the PET radiopharmaceutical $^{18}$F-Fluorodeoxyglucose (FDG PET-CT), is regarded as the imaging modality of choice for the staging and assessment of STSs [6]. FDG PET-CT provides formation about the metabolic activity of the tumor coupled to the

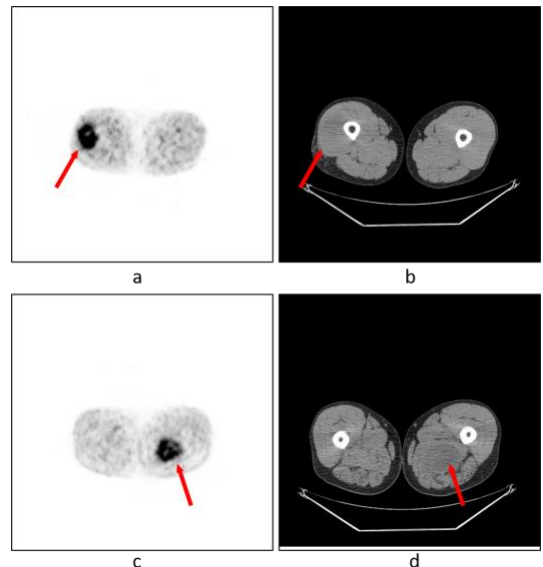

Fig. 1. Axial PET (left column) and CT (right column) of 2 patients with STSs; Red arrows indicate tumor regions. Top row (a and b) images are from a patient who did not develop DM; Bottom row (c and d) images are from a patient with DM.

¹This work was supported in part by the Australian Research Council (ARC) grants (DP200103748, IC170100022).

Y. Peng, L. Bi, D. Feng, M. Fulham, and J. Kim are with the School of Computer Science, the University of Sydney, Australia. (corresponding author emails: lei.bi@sydney.edu.au; jinman.kim@sydney.edu.au).

A. Kumar is with the School of Biomedical Engineering, the University of Sydney, Australia.

M. Fulham is also with the Department of Molecular Imaging, Royal Prince Alfred Hospital, Australia.

Y. Peng, L. Bi, A. Kumar, D. Feng, M. Fulham, and J. Kim are also with the ARC Training Centre for Innovative BioEngineering, Australia.

D. Feng is also with the Med-X Research Institute, Shanghai Jiao Tong University, China.

anatomical data that identifies where the abnormal metabolism is located. In Fig. 1, we provide two examples of STS patients with tumors in the muscles of the thighs, where although the images appear similar, only 1 of the 2 patients developed DM. Radiomics has been widely used to attempt to correlate data from medical images to disease outcomes such as the development of DM, overall survival, disease-free survival. The fundamental hypothesis in radiomics is that medical images contain information that can reflect underlying pathophysiology, and such information can be used to inform clinical decision-making [7]. Traditional radiomics methods have 3 main steps: (i) manual segmentation and annotation of regions of interest (ROIs), (ii) extracting hand-crafted (HC) radiomic features (e.g., intensity, texture, shape) and, (iii) building predictive models such as support vector machines (SVM) to correlate extracted features with the clinical outcomes [5], [8]–[19]. The performance of these methods, however, relies on a prior skillset in hand-crafting image features and tuning a large number of parameters for the predictive models.

## A. Related Work

### 1) Traditional Radiomics Studies

Early radiomics studies focused on statistical machine learning methods – support vector machine (SVM) or random forest (RF) – to associate HC radiomic features with diseases outcome. Peeken et al., demonstrated the prognostic value of hand-crafted radiomics features in STSs grading from CT and MR images respectively [8], [12]. Juntu et al. used SVM to identify the grade of malignancy of STS from magnetic resonance imaging (MRI) [13]. Corino et al. replaced SVM with a k-nearest neighbor (k-NN) for the same discrimination task [14]. Coroller et al. used univariate and multivariate analysis to prove that at least 35 radiomic features from CT images are prognostic for DM in lung cancer studies [15]. Hao et al., proposed a shall feature (consisting of outer voxels around the tumor boundary) from pre-treatment PET images to predict distant failure in non-small cell lung cancer and cervix cancer patients [16]. Li et al., developed a kernelled support tensor machine (KSTM)-based model with tumor tensors derived from pre-treatment PET and CT images to predict distant failure in early stage non-small cell lung cancer treated with stereotactic body radiation therapy [17]. Zhang et al., implemented an MR Imaging-based traditional radiomics model to predict the DM of patients with nasopharyngeal carcinoma [18]. Kwan et al., proved that hand-crafted radiomic biomarkers extracted from CT images could be used to classify DM risks for patients with nonmetastatic HPV-related Oropharyngeal Carcinoma via Cox proportional hazards model [19]. Xiong et al. evaluated the prognostic value of FDG PET radiomic features to predict local control in esophageal cancer after treatment with concurrent chemoradiotherapy [9]. Spraker et al. leveraged a penalized Cox regression model with the least absolute shrinkage and a selection operator (LASSO) algorithm to predict overall survival for STS patients on T1-weighted MR images[10]. These methods were all single-modality studies. There are few radiomics studies where multi-modality data were used. Vallières et al. showed the prognostic value of composite radiomic features, extracted from the fused PET and MR data, to identify lung metastases in STS via multivariable statistical models [5]. The same investigators used a random forest (RF) as the classifier to predict the DM in patients with head-and-neck cancer [11]. The HC features they used, from PET-CT images, included intensity solidity, skewness, grey-level co-occurrence matrix features. These methods, however, still require manual input for tumor delineation and the selection of radiomic features.

### 2) CNN-Based Radiomics Studies

In more recent reports, investigators have employed convolutional neural networks (CNNs) in radiomics [20]–[28]. CNNs extract high-level semantic information in an end-to-end manner, which reduces the need for prior knowledge in HC radiomic features definition and manual input [29]–[34]. Oakden-Rayner et al. showed that a CNN-based approach provided comparable results to conventional clinical methods in predicting longevity for patients with chronic disease [20]. Zhu et al. reported that radiomic features extracted from CNNs were superior to traditional HC features in the preoperative grading of meningiomas [21]. Hermessi et al., implemented a pre-trained AlexNet with Magnetic Resonance (MR) images in the classification of liposarcoma and leiomyosarcomas [22]. Nguyen et al., demonstrated the prognostic value of the preoperative dynamic contrast enhanced (DCE) MR images in predicting clinical lymph node metastasis of patients with breast cancer with a deep learning model [23]. Diamant et al. used a CNN-based method on pre-treatment CT scans to predict outcomes in patients with head-and-neck cancer [24]. We name this method as CNLPC for our comparison experiments setup because it consists of four main operations: convolution, non-linearity, pooling, and classification. Some investigators have coupled CNNs, as feature extractors, and traditional machine learning models [25], [26]. Lao et al. used MR images with a pre-trained CNN (trained on natural images) and then fed the CNN features into a LASSO Cox regression model to predict overall survival in patients with Glioblastoma Multiforme (GBM) [25]; GBMs are high grade primary malignant glial brain tumors. Li et al. coupled a CNN and a Fisher vector to predict the mutation status of isocitrate dehydrogenase 1 (IDH1) in patients with low grade gliomas from MR images [26], [35]. Both approaches only used a single imaging modality. There are limited studies reported on CNN-based radiomics approaches for multiple combined imaging modalities [27], [28]. Chen et al., using PET-CT in patients with head-and-neck cancer, reported a hybrid predictive model that consists of a many-objective radiomics (MOR) model and 3D CNNs to predict lymph node metastases [28]. The underlying assumption was that CNN's abstract level features and HC texture features were complementary [36], and so combining them could provide more accurate results. In our previous study, we reported a 3D CNN-based multi-modality collaborative learning (3DMCL) method that used features from PET-CT and HC radiomic features in sarcomas [27]. In both studies, using CNNs alone was sub-optimal and performance was enhanced by incorporating traditional radiomics components such as HC feature extraction. The sub-optimal performance of CNN alone was due to the extraction of the imaging features only from the last convolutional layer that contained high-level semantic information and texture information from shallow layers was neglected. Therefore, these existing CNN-based radiomics studies suffered from the limitations of traditional radiomics methods. Furthermore, they failed to fully utilize the potential of CNNs in their capability of capturing complementary information from multi-modality medical images.

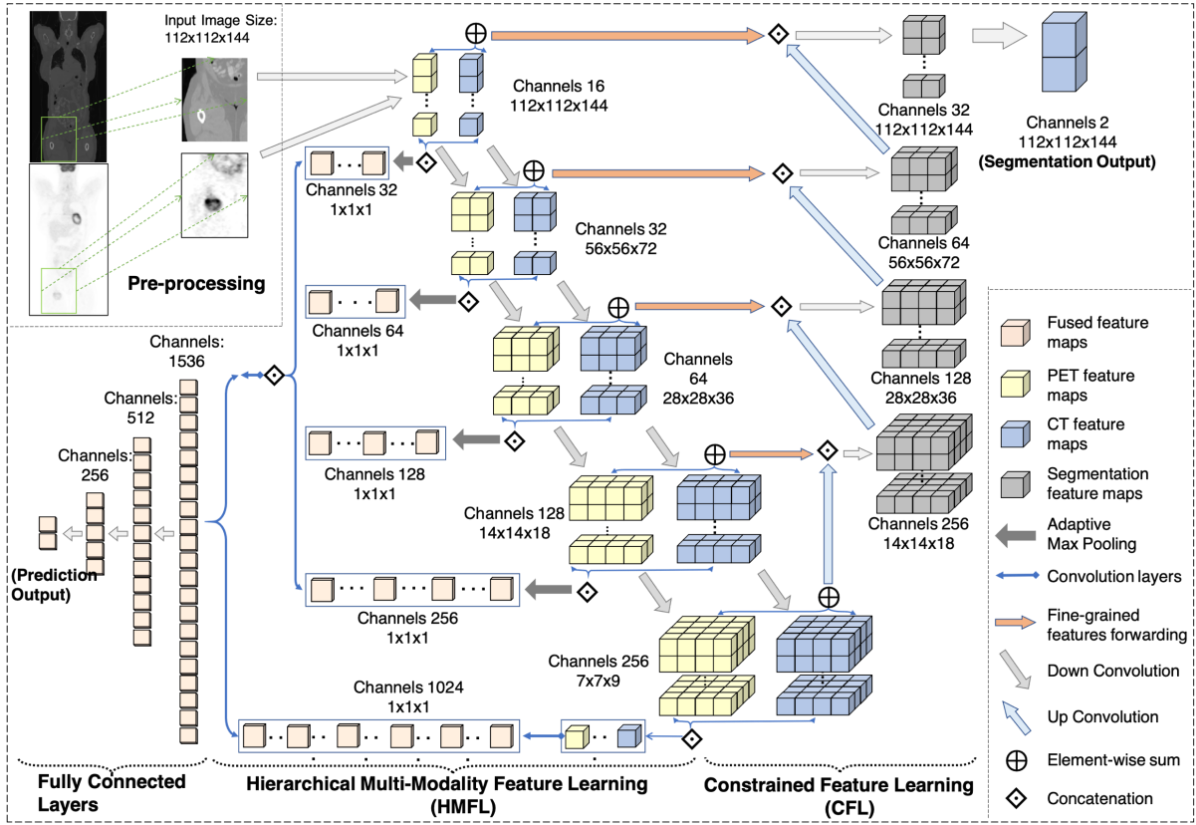

Fig. 2.  The CHMFL architecture

### B. Our Contribution

We propose a new CNN-based method to help identify the potential for DM and we refer the proposed method as constrained hierarchical multi-modality feature learning (CHMFL). Our contributions are as follows:

(i) We introduced a constrained feature learning (CFL) module to leverage functional information from PET and anatomical information from CT images to guide the learning so that it is focused on the important semantic regions (e.g., tumors). Our trained model gathered and assembled the complementary information from a multi-modality imaging modality (PET-CT) so that it automatically detected and focused on the tumor, in comparison to existing CNN-based radiomics methods that employ a single modality e.g., CT.

(ii) We designed a hierarchical multi-modality feature learning (HMFL) module to derive optimal radiomic features. Our module combined multi-modality features from different scales in an iterative manner when compared to existing radiomics methods that extracted imaging features separately from the individual imaging modalities. The hierarchical combination of features enabled a more complex and flexible fusion of PET and CT features, e.g., low-level PET texture features from a shallow layer with semantic CT features from a deeper layer.

We previously reported our preliminary work in a conference paper, using an unconstrained 3D CNN with traditional radiomic features to predict DM from FDG PET-CT [27]. This current work differs from the previous work in that we: (i) introduce a constrained feature learning module to remove the reliance on annotated ROIs during the testing stage; (ii) use a hierarchical multi-modality learning module to derive discriminative radiomics features from multiple imaging modalities and, (iii) have carried out a more exhaustive assessment of the new method in comparison with previous and more recent studies.

## II. METHODS

### A. Materials and Pre-processing

We used a public PET-CT STSs dataset from the Cancer Imaging Archive [5], [37]. This dataset has 51 patients with histologically proven, extremity primary STS. Each patient had 4 imaging modalities FDG PET, CT and T1-weighted and T2-weighted with fat-suppression (T2FS) MR scans. The gross tumor volume was manually annotated slice-by-slice on T2FS MR scans by an expert radiation oncologist and registered to PET and CT images. DM was confirmed later by biopsy or on additional imaging studies; 3 patients who developed local tumor recurrence were excluded. Hence there were 24 patients with and 24 patients without DM. The reconstructed PET image slices had a size of 96×96 with a pixel size of 3.91–5.47 $mm^2$ while CT image slice had a size of 512×512 with a pixel size of 0.98 $mm^2$. Both PET and CT had a slice thickness of 3.27 $mm$. All the PET images contained Standardized Uptake Value (SUV) data where the SUV reflects FDG uptake and is based on the dose of FDG and the body mass of the patients. Isotropic voxel resampling was applied to PET and CT images to ensure that the spatial dimensions were the same in all directions. After that, a fixed-sized bounding box of $112 \times 112 \times 144 \ mm^3$ was determined based on the size of the largest tumor in this dataset, which was used to extract the input imaging volumes. Finally, we used standardization and contrast enhancement to minimize the influence of both high and low-frequency noise [38].

### B. Overview of the Proposed Method

In Fig. 2, we outlined our CHMFL architecture. The volumetric PET and CT images were pre-processed and then fed separately into two identical branches. Each branch has multiple downsampling convolutional layers for feature extraction (as shown within the yellow PET and blue CT feature maps in Fig. 2). The feature maps derived after each convolutional layer were adaptively pooled and then concatenated into a single feature vector to facilitate hierarchical multi-modality feature learning (HMFL). The constrained feature learning (CFL) module used several upsampling convolutional layers to guide the network to focus on the important regions (e.g., the tumor). This process also incorporated the fine-grained features forwarded from HMFL module at each level. Finally, the derived multi-modality PET-CT features (as shown in the left lower part of Fig. 2) were fed into three fully connected layers for distant metastases (DM) prediction.

### C. Constrained Feature Learning (CFL) Module

Our CFL module was designed to guide the learning process to focus on semantically important regions at both the training and inference stages. This was achieved by gathering and assembling the complementary information from multi-modality PET-CT images to obtain a 2-channel volumetric segmentation output. We used 4 transposed convolutional blocks to expand the spatial support from the feature maps at a lower scale for upsampling. These upsampling blocks at different levels shared similar structures (see CFL module in supplementary Table SI for details). Meanwhile, the multi-modality PET-CT features extracted from the HMFL module were forwarded to the upsampling blocks by horizontal connections (see Fig. 2). In this way, we gathered fine-grained detail for tumor contour prediction that would be otherwise lost in the downsampling path. In turn, tumor regions were emphasized in the HMFL module by the backpropagation process. Moreover, in order to avoid the vanishing gradient problem with network deepening, a residual learning was formulated after the concatenation of forwarded PET-CT features and the corresponding upsampled feature maps at each level: the concatenated feature map was processed through several convolutional layers and non-linearities, then added to the output of the last non-linearity within the residual learning.

During the training stage, two loss functions were employed for different tasks. A pixel-wise cross-entropy loss was used to compare the predicted segmentation output with the ground truth tumor annotation. Another cross-entropy loss was used for DM prediction. Given a weight $w$ for our CFL module $0 \le w \le 1$, the total loss $L$ was defined as follows:

$$L = -(1-w) * \sum_{m=1}^{M=2} p_{1,m} \log q_{1,m} - w * \sum_{n=1}^{N} p_{2,n} \log q_{2,n} \qquad (1)$$

where $p_{1,m}$ represents the target probability of DM, $q_{1,m}$ (the output of this network) represents the predicted probability of developing DM, and $M$ denotes the number of output neurons generated by the last fully connected layer in this network, $q_{2,n} \in Q$ is the predicted binary segmentation volume, $p_{2,n} \in P$ is the ground-truth binary annotation image and $N$ denotes for the total number of image voxels. $w_i$ is a weight to balance the two losses.

### D. Hierarchical Multi-modality Feature Learning (HMFL) Module

We used 5 convolutional blocks for multi-modality image feature extraction (more details of HMFL module are provided in supplementary Table SI). PET and CT images were processed separately by the identical PET and CT branches. Within each convolutional block, the output feature map of the 3D convolutional layer was defined as:

$$F = W * X + b \qquad (2)$$

where $X$ is the input to the convolution layer, * is the convolution operation, $W$ denotes for the learned weights, and $b$ is the learned bias. A batch normalization layer and a non-linear activation function ELU were also added. By performing a 3D convolution with a kernel size of $(I, J, K)$, the value at location $(x, y, z)$ of the feature map $F$ was determined from its neighborhood:

$$F(x, y, z) = \sum_i \sum_j \sum_k W(i, j, k) * X(x+i, y+j, z+k) \quad (3)$$

with $-\lfloor \frac{I}{2} \rfloor \le i \le \lfloor \frac{I}{2} \rfloor$, $-\lfloor \frac{J}{2} \rfloor \le j \le \lfloor \frac{J}{2} \rfloor$, $-\lfloor \frac{K}{2} \rfloor \le k \le \lfloor \frac{K}{2} \rfloor$.

For hierarchical multi-modality feature learning (HMFL), we firstly concatenated PET and CT feature maps to include multi-modality context information at each scale of the convolutional layers. After concatenation, an adaptive pooling layer was used to project the fused feature map into a single vector. This combination of feature maps from different scales could obtain both diverse texture details from shallow layers and high-level semantic layers, which can be defined as:

$$F_{fusion} = F(\cup_{l=1}^{L=4} APL(F_{pet}^l \otimes F_{ct}^l)) \otimes F(F_{pet}^5 \otimes F_{ct}^5) \qquad (4)$$

Where $APL$ denotes for the adaptive max pooling layer, and $L$ is the number of convolutional layers for multi-modal PET-

TABLE I. CLASSIFICATION PERFORMANCE COMPARISONS WITH EXISTING RADIOMICS METHODS

| Method | Evaluation Metrics | | | | | |
|--------|------|------|------|------|------|------|
| | *Acc.* | *Sen.* | *Spe.* | *Pre.* | *F1.* | *AUC* |
| HC + RF | 0.750* | 0.792* | 0.708* | 0.731* | 0.760* | 0.726* |
| CNLPC | 0.729* | 0.792* | 0.667* | 0.703* | 0.745* | 0.783* |
| MOR + 3D CNNs | 0.729* | 0.750* | 0.780* | 0.720* | 0.750* | 0.793* |
| 3DMCL | 0.854* | 0.917* | 0.797* | 0.813* | **0.862*** | 0.854* |
| Mask + HMFL (Ours) | 0.813 | 0.750 | **0.875** | 0.857 | 0.800 | 0.769 |
| CFL (Ours) | 0.813 | **0.958** | 0.667 | 0.742 | 0.836 | 0.852 |
| CHMFL (Ours) | **0.854** | 0.875 | 0.833 | 0.840 | 0.857 | **0.873** |

TABLE II. CLASSIFICATION PERFORMANCE COMPARISONS WITH METHODS USING DIFFERENT MODAL IMAGE AND CONVOLUTIONAL LAYERS

| Method | Evaluation Metrics | | | | | |
|--------|------|------|------|------|------|------|
| | *Acc.* | *Sen.* | *Spe.* | *Pre.* | *F1.* | *AUC* |
| 2D-CNN-CT | 0.583* | 0.708* | 0.458* | 0.567* | 0.630* | 0.503* |
| 2D-CNN-PET | 0.729* | 0.542* | 0.917* | 0.867* | 0.667* | 0.656* |
| 2D-CNN-PET-CT | 0.729* | 0.792* | 0.667* | 0.703* | 0.745* | 0.698* |
| 3D-CNN-CT | 0.667* | 0.667* | 0.667* | 0.667* | 0.667* | 0.684* |
| 3D-CNN-PET | 0.771* | 0.750* | 0.792* | 0.783* | 0.766* | 0.734* |
| 3D-CNN-PET-CT | 0.792* | 0.792* | 0.792* | 0.792* | 0.792* | 0.773* |
| Mask + HMFL (Ours) | 0.813 | 0.750 | **0.875** | **0.857** | 0.800 | 0.769 |
| CFL (Ours) | 0.813 | **0.958** | 0.667 | 0.742 | 0.836 | 0.852 |
| CHMFL (Ours) | **0.854** | 0.875 | 0.833 | 0.840 | **0.857** | **0.873** |

*: $p < 0.05$, in comparison to our proposed CHMFL method derived from an unpaired student's t-test.

CT feature extraction, and ($\otimes$) represents the concatenation operation.

Multi-modality feature maps at each scale were concatenated into a single fully connected layer and processed with additional two fully connected layers. ReLU layers and dropout layers with a probability of 0.5 were added after each fully connected layer to reduce overfitting.

### E. Implementation Details

Our method was implemented with PyTorch [39] and ran on an 11GB NVIDIA GeForce GTX 1080Ti GPU. The learning rate was set to 0.0001 and the batch size was set to 1. Our model was initialized using the approach presented in He et al. [40], and adaptive-moment-estimation (Adam) [41] was used for network optimization. Training was terminated when no further changes in the total loss. In our method, the total loss was generally stable after 200 epochs and our CHMFL model took approximately five hours to fine-tune with. In addition, our model took around 10 seconds to inference 8 patients; this time is similar to the existing 3D based CNN models.

### F. Experimental Setup

We performed the following experiments where we: (a) compared the proposed method with the state-of-the-art radiomics methods; (b) compared the performance of using PET or CT images only; (c) compared the performance of using 2D CNNs or 3D CNNs and, (d) analyzed the individual components of the proposed method: (i) CFL - our proposed method without HMFL module, used PET and CT images as input; (ii) Mask + HMFL - our proposed method without CFL module, used PET, CT and tumor label images as input (2-channel PET-label image and 2-channel CT-label image). The state-of-the-art methods used in our comparisons were those mentioned in the related works, and they can be divided into three categories: (i) traditional radiomics method - HC+RF [5]; (ii) CNN-based radiomics method - CNLPC [24] and, (iii) hybrid methods that combine CNNs with traditional radiomic components: 3DMCL [27] and MOR+3D CNNs [28]. We used a hold out 6-fold cross-validation approach for our method and the methods that we compared it to. The 48 PET-CT data were randomly divided into 6 equal-sized subsets and each subset had 8 PET-CT images. For each fold, 5 subsets were used to train the network and the remaining subset were used for testing. We repeated this process 6 times to assess the 48 PET-CT images. The results that we present in Section III are the mean value across all 6 folds. Six established evaluation metrics were adopted including accuracy (acc.), sensitivity (sen.), specificity (spe.), precision (pre.), F1 score (F1) and, area under the receiver-operating characteristic curve (AUC). For all experimental comparisons with our proposed CHMFL method, we computed the *p*-value with an unpaired student's *t*-test.

### III. RESULTS

Our CHMFL achieved the overall best DM prediction performance with the highest accuracy (0.854) and AUC (0.873) - see Table I and Fig. 3; Our CHMFL's F1 score (0.857), specificity (0.833) and precision (0.840) ranked at the second place, and sensitivity (0.875) is the third-best. In addition, our CFL module alone obtained the highest sensitivity (0.958), and our Mask + HMFL obtained the highest specificity (0.875) and precision (0.857).

When compared with methods using 2D or 3D CNNs with different modality imaging data (e.g., PET, CT, PET-CT) - see Table II and Fig. 4, our CHMFL method outperformed all the comparison CNNs based methods regardless of imaging modality and the kernel dimension of CNN. 3D CNNs performed better than those using 2D CNNs. Methods based on PET images outperformed methods based on CT images.

We evaluated how the CFL module's weight *w* affected the performance of CHMFL over three key evaluation metrics (e.g.,

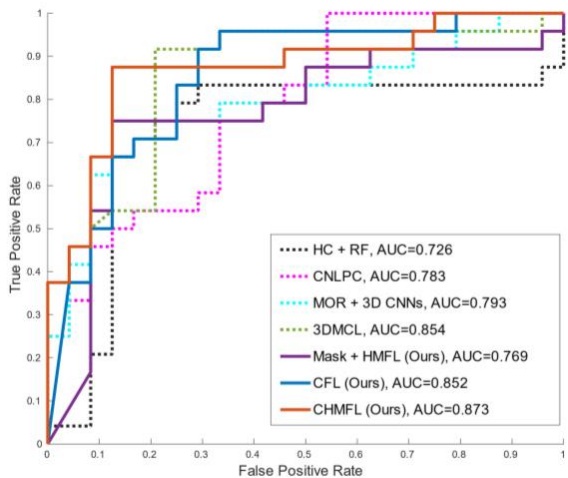

Fig. 3. Classification performance (measured in receiver operating characteristic (ROC) curve) of our CHMFL in comparison to other existing radiomics methods.

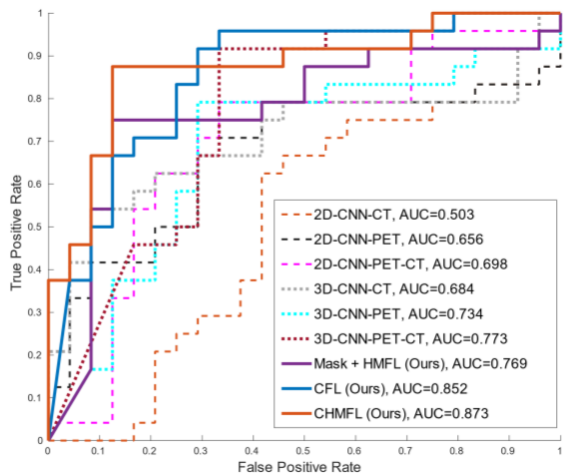

Fig. 4. Classification performance (measured in ROC) of our CHMFL in comparison to methods using different modal image and convolutional layers.

accuracy, sensitivity and specificity). The result in Fig. 5 suggests that the best performance was achieved when the weight was 0.5. We also evaluated the segmentation results produced by the CFL module (see supplementary section SIV for details).

Two example PET-CT studies are shown in Fig. 6 with corresponding visualization results of the extracted feature map with respect to three existing radiomics methods that outperformed other comparison methods except for our CFL and CHMFL methods.

The discriminative ability of both the 3DMCL and our CFL and CHMFL methods is depicted in Fig. 7, via t-distributed stochastic neighborhood embedding (t-SNE) [42] visualization. t-SNE is an unsupervised, non-linear technique primarily used for visualizing high-dimensional image features in a two or three-dimensional space, which allows for exploring the relationship of the extracted features.

## IV. DISCUSSION

Our main findings are that: (i) our CFL module automatically identified the tumor; (ii) our HMFL module derived multi-modality PET-CT image features; (iii) our method improved upon current single-modality methods and, (iv) our CHMFL outperformed state-of-the-art radiomics methods.

### A. CFL Module Analysis

Our CFL automatically focused on ROIs that were semantically more important to the STSs DM predictions. When compared to 3D-CNN-PET-CT (Fig. 6 (f)), both our CFL and CHMFL methods with the CFL module were able to correctly focus on tumor regions in the derived feature maps (Fig. 6(c) and Fig. 6(d)). In contrast, 3D-CNN-PET-CT falsely concentrated on many normal uptake regions, which resulted in a >10% decrease in AUC value and > 4% decrease in both accuracy and F1 score (shown in Table II). Moreover, when compared to existing methods that segmented tumor region before feature extraction and outcome prediction, such as 3DMCL (the state-of-the-art method on STS DM prediction, Fig. 6 (e)), both our CHMFL and CFL methods could accurately identify the entire sarcoma with more details.

Although Mask + HMFL was forced to focus on the tumor region by incorporating an extra channel of tumor label image as input, there were still some false positive regions, e.g., as in the bottom case of Fig. 6 (e) when there were similar tissues around the tumor. Without our CFL module, the Mask + HMFL method had a tendency of not predicting DM due to concentrating on more regions other than tumors. Although this contributed to a 4% increase in specificity and 1% increase in precision, there are >10% decrease in both AUC score and sensitivity when compared with our proposed CHMFL (shown in Table II). We also noted that automatically constraining the learning process to extract feature maps only from the tumor region can obtain more information that can potentially reflect underlying pathophysiology, such as the heterogeneity of STS, which is an important prognostic factor of DM development [43]. In addition, such an automated process removes the reliance on accurate manual tumor delineation during the inference stage while obtaining better overall performance.

### B. HMFL Module Analysis

The inclusion of HMFL in our method further improved the performance of the CFL. Most existing CNNs based radiomics methods, including 3DMCL, CNLPC and MOR+3DCNNs, only leverage high-level features extracted from the last convolutional layer in their model, and therefore inherently disregarded the complementary PET and CT image features at the lower level of the network. In contrast, our method iteratively and hierarchically fused the multi-modality PET and CT image features across the different image scales, which enabled more flexible and complex multi-modality information fusion. As an example, the feature map derived from our CHMFL method (Fig. 6 (c)) captured more details inside the tumor and better predicted the tumor contour when compared with our CFL method (Fig. 6 (d)).

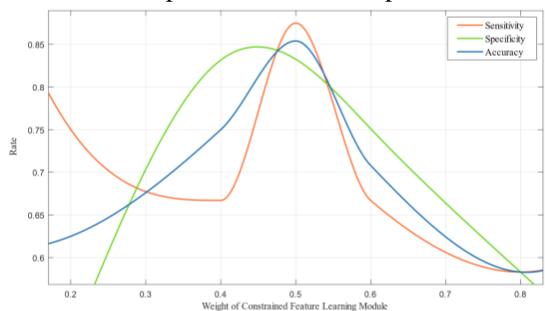

Fig. 5. Analysis of the weight used in the CFL module.

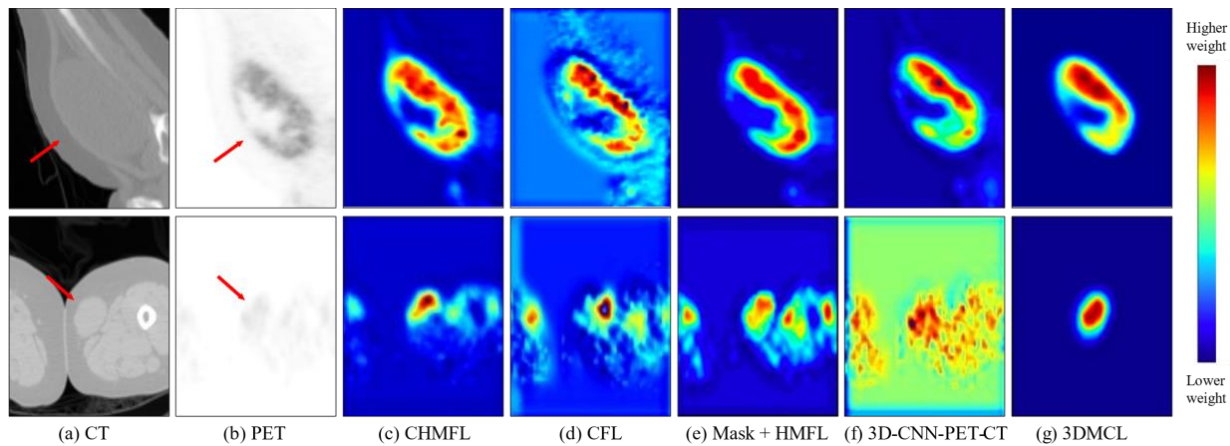

Fig. 6. The original CT (a) and PET (b) images, and corresponding feature map visualization results of CHMFL (c), CFL (d) and other radiomics methods: (e) Mask + HMFL, (f) 3D-CNN-PET-CT, and (g) 3DMCL. Images have been cropped to the tumor ROI and the red arrows indicate the tumor regions. Blue color in these feature map visualization results indicates low weight, whereas yellow and red indicate higher weight.

### C. Evaluation of CNN-Based Methods with Different Image Modalities and Different Convolutional Layers

There was a marked difference in performance between PET-CT CNNs and CNNs with PET alone or CT alone. Further, PET-based methods outperformed CT-based methods. This was expected since PET images provided metabolism information of tumors, while CT can only provide the anatomical information, and tumor regions are not always visible in CT (as exemplified in Fig. 1). The relatively lower performance of 2D CNNs, when compared to 3D CNNs counterparts is attributed to the fact that volumetric image features derived from 3D CNNs are better to discriminate the spatial information within the tumor that is associated with the DM development, e.g., volumetric tumor shape and size [44]. In contrast, 2D CNNs based methods (e.g., 2D-CNN-CT and 2D-CNN-PET) have limited representation capability of tumor characteristics in two dimensions with few axial slices. Therefore, it would be better to incorporate 3D CNNs with multi-modality imaging data when the computational power is available, which allows to achieve better performance (as shown in Table II).

### D. Comparison of CHMFL with Existing Methods

Our CHMFL method obtained the best overall performance when compared with the existing radiomics methods. HC+RF method achieved competitive performance over all the evaluation metrics except the AUC score when compared with CNLPC and MOR+3DCNNs. Unfortunately, the performance of HC+RF was reliant on effective feature handcrafting and tuning a large number of parameters, which may limit its generalizability to different datasets. The performance improvement from 3DMCL to CNLPC and MOR+3DCNNs was likely due to the use of multi-modality PET and CT images providing complementary information. When compared to the second-best performing method 3DMCL, our method achieved much higher specificity (as shown in Table I). 3DMCL is reliant on using single-level image features for prediction, which results in 3DMCL overfits to the positive prediction of DM. which were unable to discriminate the tumors. As exemplified in Fig. 7 our CHMFL had greater separability between the patients with/without DM than both 3DMCL and 3D-CNN-PET-CT, where only a few cases were not properly separated.

### E. Limitations and Future Work

Our focus in the current study was to investigate the prediction of distant tumor spread (metastatic disease) in patients with STSs from PET-CT images. Predicting the presence of distant metastases (DM) as a binary classification is an abstraction of a time to event prediction problem (i.e., estimating the point at which an event occurs). The time to event problem is a more complicated modelling challenge than binary classification and may require different methodological approaches. In the public dataset we used, all the patients have a 7-year follow-up period for outcome observation and DM was generally confirmed within 4 years after diagnosis of primary STS; this was appropriate for binary classification. The public dataset is small (n=51) and thus there was no separate held-out data used only for testing. We reported only the mean results across all validation experiments. The results may be different with a held-out cohort in a much larger dataset. We are actively working on characterizing and annotating a much larger soft tissue sarcoma dataset. Moreover, we have not generalized our results to other tumor types or where other imaging modalities are employed. In future work we intend to evaluate our approach in non-small cell lung cancer and the lymphomas, using PET-CT, and also include other parameters such as local tumor recurrences and long-term survival. We would also like to adapt our approach to other imaging modalities such as PET-Magnetic Resonance (PET-MR) and parametric MRI imaging.

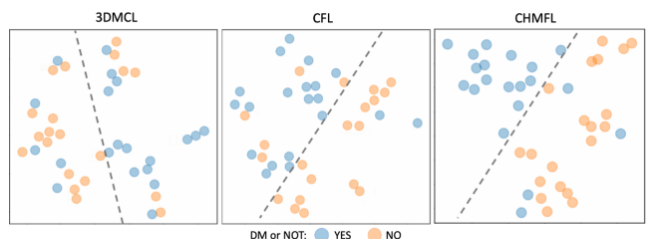

Fig. 7. t-SNE visualization result of 3DMCL, CFL and CHMFL methods. A dashed line is added to demonstrate how the features are separated.

## V. Conclusions

We proposed a constrained hierarchical multi-modality feature learning method for predicting the development of distant metastases. Our results with a public dataset of soft tissue sarcomas showed that our method was capable to better identify PET-CT radiomics features in primary tumors that were associated with the development of DM, when compared to the state-of-the-art radiomics methods.

# Supplementary Materials for Manuscript "Predicting Distant Metastases in Soft-Tissue Sarcomas from PET-CT scans using Constrained Hierarchical Multi-Modality Feature Learning"

## SI. THE TABLE CONTAINING ARCHITECTURE DETAILS

Table SI illustrates the network architecture details used in our proposed hierarchical multi-modality feature learning (HMFL) and constrained feature learning (CFL) modules.

TABLE SI. NETWORK ARCHITECTURE USED IN THE HMFL AND CFL MODULE

| Layers | Details (kernel size, stride, padding, …) | Output Size (batch size, channel number, …) |
|---|---|---|
| **HMFL Module** | | |
| Input Transition | Conv3d (5×5×5, 1, 2); BatchNorm; ELU | 1×16×112×112×144 |
| Down_Conv_1 | Conv3d (2×2×2, 2, 0); BatchNorm; ELU; | 1×32×56×56×72 |
| Down_Conv_2 | Conv3d (2×2×2, 2, 0); BatchNorm; ELU; | 1×64×28×28×36 |
| Down_Conv_3 | Conv3d (2×2×2, 2, 0); BatchNorm; ELU; | 1×128×14×14×18 |
| Down_Conv_4 | Conv3d (2×2×2, 2, 0); BatchNorm; ELU; | 1×256×7×7×9 |
| **CFL Module** | | |
| Up_Conv_1 | ConvTranspose3d (2×2×2, 2, 0); BatchNorm; ELU | 1×128×14×14×18 |
| Up_Conv_2 | ConvTranspose3d (2×2×2, 2, 0); BatchNorm; ELU | 1×64×28×28×36 |
| Up_Conv_3 | ConvTranspose3d (2×2×2, 2, 0); BatchNorm; ELU | 1×32×56×56×72 |
| Up_Conv_4 | ConvTranspose3d (2×2×2, 2, 0); BatchNorm; ELU | 1×16×112×112×144 |
| Output Transition | Conv3d (5×5×5, 1, 2); BatchNorm; ELU; Conv3d (1×1×1, 1, 0); softmax | 1×2×112×112×144 |

The confusion matrix allows to compare non-averaged results for 2-class problem. We have included the confusion matrixes of the top-4 methods from our experiments for additional comparisons (see Fig. S1 below). When compared with the existing radiomics methods, our CHMFL method obtained overall better performance with the lowest false positive rate (8.3% out of 48 patients) while maintained the highest accuracy (85.4%).

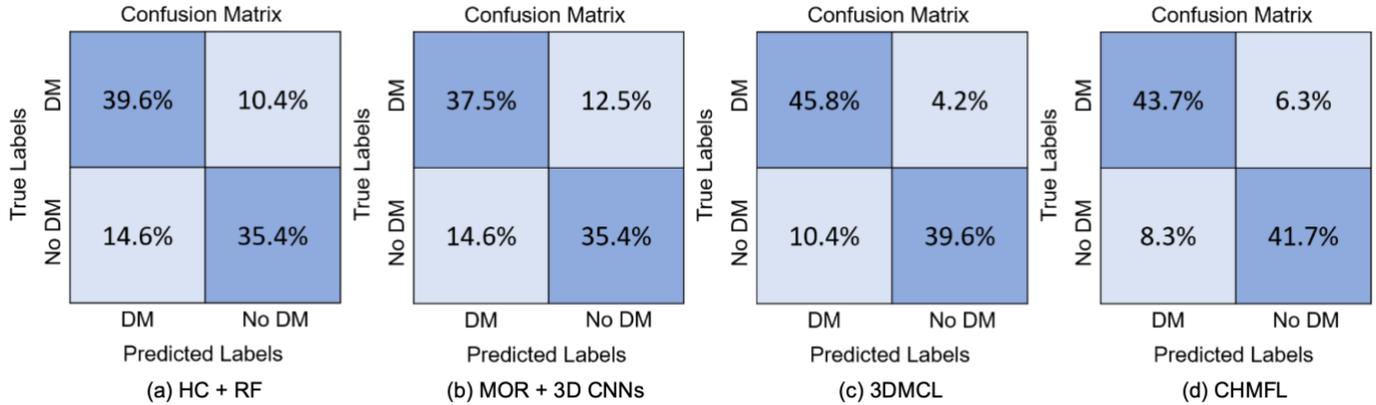

Fig. S1.  The Confusion Matrix of (a) HC + RF, (b) MOR + 3D CNNs, (c) 3DMCL and (d) CHMFL.

## SIII. RESULTS OF TRANSFER LEARNING BASED METHODS

The current transfer learning based methods focus on using pre-trained models trained from large natural image datasets e.g., ImageNet, and then fine-tune the pre-trained model on the smaller target medical imaging data. However, our proposed CHMFL method is based on 3D, which is not applicable for transfer learning from the 2D image pre-trained models. Therefore we reimplemented our method in 2D and used ResNet model as the backbone to facilitate the transfer learning process. We performed additional experiments to compare the performance of DM prediction in STSs patients with and without transfer learning.

Our results show that the baseline methods (ResNet) and our CHMFL benefited from transfer learning (see Table SII below). The results also indicate that the CHMFL method, in 2D with transfer learning, consistently outperformed the baseline methods with transfer learning. In addition, CHMFL in 3D further improved the performance over the five evaluation metrics and had the 2nd-best sensitivity of 0.875. Although the CHMFL + ResNet18 (pretrained) obtained the highest sensitivity (0.917), the relatively low specificity and accuracy illustrated a tendency to overfitting in the positive prediction of DM, which means that CHMFL + ResNet18 (pretrained) may have limited generalizability.

TABLE SII. CLASSIFICATION PERFORMANCE COMPARISONS WITH TRANSFER LEARNING METHODS

| Method | Evaluation Metrics | | | | | |
| --- | --- | --- | --- | --- | --- | --- |
| | Acc. | Sen. | Spe. | Pre. | F1. | AUC |
| 2D-CNN-PET-CT | 0.729 | 0.792 | 0.667 | 0.703 | 0.745 | 0.698 |
| ResNet18-PET-CT | 0.729 | 0.708 | 0.750 | 0.739 | 0.723 | 0.695 |
| ResNet18-PET-CT (pretrained) | 0.792 | 0.875 | 0.708 | 0.750 | 0.808 | 0.705 |
| ResNet34-PET-CT | 0.708 | 0.667 | 0.750 | 0.727 | 0.696 | 0.683 |
| ResNet34-PET-CT (pretrained) | 0.750 | 0.708 | 0.792 | 0.773 | 0.739 | 0.717 |
| ResNet50-PET-CT | 0.729 | 0.667 | 0.792 | 0.762 | 0.711 | 0.653 |
| ResNet50-PET-CT (pretrained) | 0.750 | 0.833 | 0.667 | 0.714 | 0.769 | 0.658 |
| CHMFL+ ResNet18 | 0.750 | 0.792 | 0.708 | 0.731 | 0.760 | 0.715 |
| CHMFL+ ResNet18 (pretrained) | 0.792 | **0.917** | 0.667 | 0.733 | 0.815 | 0.741 |
| CHMFL+ ResNet34 | 0.729 | 0.750 | 0.708 | 0.720 | 0.735 | 0.701 |
| CHMFL+ ResNet34 (pretrained) | 0.771 | 0.833 | 0.708 | 0.741 | 0.784 | 0.726 |
| CHMFL+ ResNet50 | 0.729 | 0.625 | 0.833 | 0.790 | 0.698 | 0.710 |
| CHMFL+ ResNet50 (pretrained) | 0.771 | 0.875 | 0.667 | 0.724 | 0.793 | 0.754 |
| CHMFL (Ours) | **0.854** | 0.875 | **0.833** | **0.840** | **0.857** | **0.873** |

## SIII. RESULTS OF GROUP NORMALIZATION BASED METHODS

We have conducted additional experiments by replacing the batch normalization (BN) approach used in our CHMFL method with a group normalization (GN) approach [1]. We set GN channels as 4, 8 and 16, which are denoted as CHMFL + GN_c4, CHMFL + GN_c8 and CHMFL + GN_c16. The experimental results are shown in the Supplementary Table SIII and our method with the BN approach had consistently better performance than the GN-based approaches.

TABLE SIII. CLASSIFICATION PERFORMANCE COMPARISONS WITH GROUP NORMALIZATION METHODS

| Method | Evaluation Metrics | | | | | |
| --- | --- | --- | --- | --- | --- | --- |
| | Acc. | Sen. | Spe. | Pre. | F1. | AUC |
| CHMFL + GN_c4 | 0.729 | 0.583 | **0.875** | 0.824 | 0.683 | 0.706 |
| CHMFL + GN_c8 | 0.771 | 0.792 | 0.750 | 0.760 | 0.775 | 0.787 |
| CHMFL + GN_c16 | 0.833 | 0.875 | 0.792 | 0.808 | 0.840 | 0.820 |
| CHMFL (Ours) | **0.854** | **0.875** | 0.833 | **0.840** | **0.857** | **0.873** |

# SIV. Segmentation Results of Our Proposed CHMFL Method

We compared the segmentation performance of our proposed methods with V-net (a commonly used segmentation method for volumetric medical images) [2]. Five established evaluation metrics were adopted for the segmentation evaluation, including dice similarity coefficient (DSC), accuracy (acc.), Jaccard score (jac.), sensitivity (sen.), and specificity (spe.).

The experimental results showed that our CHMFL obtained overall better segmentation performance (see Table SIV and Fig. S2). Although V-net obtained the highest specificity (0.999), this is resulted from the less-segmentation of tumor region (see Fig. S2). The higher sensitivity (0.991) of CFL method was attributed to the over-segmentation (see Fig. S2), which resulted in a lower voxel-level accuracy (0.960). Whereas CHMFL with hierarchical multi-modality feature learning module was able to capture more nuanced morphological details that are more critical in segmentation.

TABLE SIV. Segmentation Results of Our Proposed CHMFL Methods

| Method | Evaluation Metrics | | | | |
|---|---|---|---|---|---|
| | *DSC* | *Jac.* | *Acc.* | *Sen.* | *Spe.* |
| V-net | 0.716 | 0.558 | 0.987 | 0.561 | **0.999** |
| CFL (Ours) | 0.707 | 0.547 | 0.960 | **0.991** | 0.941 |
| CHMFL (Ours) | **0.722** | **0.565** | **0.988** | 0.898 | 0.989 |

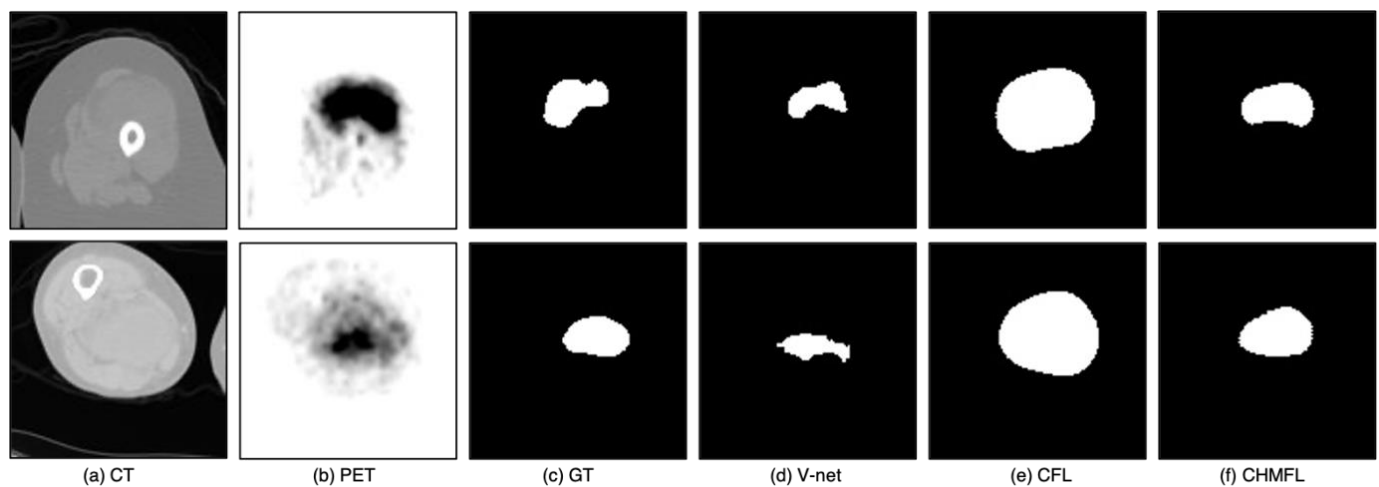

(a) CT      (b) PET      (c) GT      (d) V-net      (e) CFL      (f) CHMFL

Fig. S2. Segmentation results from two examples (top row; bottom row) of STSs. The images are transaxial views of: (a) input CT (after pre-processing); (b) input PET; (c) ground truth (GT) annotation; (d), (e) and (f) segmentation results from V-net, CFL and CHMFL methods.